\documentstyle[epsf,latexsym]{l-aa}
\newcommand{\be}{\begin{equation}}
\newcommand{\ee}{\end{equation}}
\newcommand{\hr}{H\,{\sc ii} region}
\newcommand{\hrs}{H\,{\sc ii} regions}
\newcommand{\sii}{[\rm S\,{\sc ii}]}
\newcommand{\siii}{[\rm S\,{\sc iii}]}
\newcommand{\oii}{[\rm O\,{\sc ii}]}
\newcommand{\oiii}{[\rm O\,{\sc iii}]}
\newcommand{\hbeta}{{\rm H}\beta}
\newcommand{\hii}{\rm H\,{\sc ii}}
\newcommand{\spunk}{\rm S_{23}}

\begin{document}
\thesaurus{3(09.08.1; 11.01.1; 11.09.1 NGC 300; 11.19.2)}
\title{A sulphur abundance study of NGC 300 by an empirical calibration method
\thanks{Based on observations obtained with the Danish 1.5-m telescope at ESO,
La Silla, Chile}}
\author{T. Christensen\inst{1,2}\and L. Petersen\inst{1} \and
 P. Gammelgaard\inst{1}}
\offprints{tina@obs.aau.dk, larsp@obs.aau.dk}
\institute{Institut for Fysik og Astronomi, Aarhus Universitet, Ny
 Munkegade, DK-8000 \AA rhus C, Denmark \and Nordic Optical Telescope, Apartado
474, E-38700 Sta. Cruz de La Palma, Spain}
\date{Received \hskip 2cm; Accepted}
\maketitle
\markboth{Christensen et al.: A Sulphur abundance study of NGC 300 by an 
empirical calibration method}{Sulphur abundance in NGC 300}

\begin{abstract}

We propose an empirical sulphur abundance determination method based
on the strong sulphur emission lines, \sii ~$\lambda \lambda$6716, 6731 and  
\siii ~$\lambda \lambda$9069, 9531 in \hrs . 
From a compilation of literature data we have made a calibration of 
sulphur abundance versus (\sii ~+ \siii )/$\hbeta$, similar to what has been
widely used for the more easily observable iso-electronic element oxygen.
This enables abundance determinations in extragalactic \hrs ~ without 
measurements of weak temperature sensitive lines for use in model calculations.

As a first application of the empirical calibration 15 spectra covering the 
wavelength range 3650--10000 \AA ~have been obtained of \hrs ~at
various galactocentric distances in the spiral galaxy NGC 300, 
by employing specially designed multiaperture slitmasks which allow
simultaneous observations of several \hrs . 
Sulphur and oxygen abundances are determined
and an oxygen gradient in agreement with previously published work
is found as well as a somewhat steeper sulphur abundance gradient.
This results in a slight decrease of log S/O with increasing radius.

The quite low values of S/O seem to confirm the trend demonstrated
by D\'\i az et al. (1991) of a decrease in log S/O at high
metallicities, but still more data are needed before any definite
conclusions can be drawn. We find as expected that the radiation 
softness parameter $\eta$ decreases with increasing radius and
the slope is comparable to what is found for spiral galaxies of
similar metallicity.
\end{abstract}

\section{Introduction}
Elemental abundance studies in spiral galaxies have largely been concerned
with the abundance of oxygen and the presence of radial O-gradients with
negative slopes have been demonstrated for more than a decade. Initiated by the
analyses and line ratio calibrations of Edmunds \& Pagel (1984) and McCall et 
al. (1985) large observational samples have recently been compiled by several
groups (Vila-Costas \& Edmunds 1992; Oey \& Kennicutt 1993; Zaritsky et al.
1994, hereafter ZKH), of which the latter collates much of the existing
published and their own spectroscopic data on a total of 577 \hrs~ in 39
spirals.

The present data available through these surveys are of such proportions, that
they can address the correlation of oxygen abundance with local and macroscopic 
parameters of spiral galaxies, which can be used to constrain galaxy evolution
models.
ZKH define a characteristic oxygen abundance and find that it correlates with
galaxy hubble type and luminosity. No clear correlation of oxygen abundance
gradients, when expressed as dex/isophotal radius, with Hubble type is seen,
although there are indications that intermediate-type spiral galaxies have the
steepest gradients and that the slope is flatter in barred than unbarred
spirals.

Much less attention has been paid to the sulphur content of spiral galaxies,
probably because the strong lines of the ionization stage S$^{2+}$ is
situated in the near-IR region at $\lambda =$ 9069 and 9531
\AA{}, where detector sensibility is low and many atmospheric emission lines are
present, although it can give important clues on elemental abundance and the 
softness parameter as defined in Section 5.1, which was shown by ZKH to
increase with decreasing radius, etc.

Garnett (1989) investigates sulphur abundance in 13 extragalactic \hii 
~regions and summing up on his own and previous work concludes that there is
only very weak if any  evidence for gradients in S/O 
in spiral galaxies. Neither is there a systematic variation of S/O with
oxygen abundance even though it might be expected from chemical evolution
models. In this work we will put more emphasis on the sulphur abundance in
spiral galaxies by developing an empirical calibration of sulphur line ratios
and abundance.

It has become evident from the studies cited above, that conclusions on
abundance
variation over a galaxy disk drawn on data from less than five \hrs ~can be very
misleading
and even turn out with the wrong sign of the gradient. Therefore more
than 10 observed regions per galaxy with good radial coverage is desirable. 
Ryder (1995), who added oxygen data on 6 new galaxies with findings in support
of ZKH, has discussed how \hrs~ can be observed in a time efficient
manner with narrow band filters and multifiber or multislit spectroscopy and 
pointed
out some of the drawbacks of the first two. When broad wavelength coverage of
\hrs~ in nearby galaxies is the objective, multislit spectroscopy might be
prefered, since most regions extend over many arcseconds and may be closely
situated.
Hence, they are not efficiently observed using
multifiber spectrographs designed for point-like objects.

NGC 300 is a large, nearby spiral galaxy
with many prominent \hrs. At the modest distance of the Sculptor group
of galaxies, about 1.2 Mpc, the Scd galaxy extends over 20\arcmin~ and is a
perfect object for multislit spectroscopy covering wide fields. By this 
technique using specially designed multiaperture plates the number of
\hrs{} that can be observed simultaneously is increased as compared to
conventional long-slit spectroscopy.

\section{Line ratios as abundance indicators}

\noindent
Abundances from \hii ~region emission lines are most accurately determined
when temperature sensitive lines are measured, since the derived temperature
enables modeling of the region. The required lines such as [\rm O\,{\sc iii}]
$\lambda$4363 are, however, quite weak and often not detectable at all.
In their absence one has to depend on empirical relationships between the
desired abundance and a combination of strong emission lines. Pagel et al.
(1979) suggest $\rm R_{23}$ = (\oii ~$\lambda \lambda$3726, 3729 
+ \oiii ~$\lambda \lambda$4959, 5007)/$\hbeta$ as an improved oxygen 
abundance indicator as compared to earlier used line ratios. 

A calibration of log($\rm R_{23}$) against oxygen abundance by Edmunds \& 
Pagel (1984) has since been widely used for abundance determinations in 
regions where the temperature sensitive lines are unreachable.
The fit is ambiguous, however, consisting of two branches, and care must 
be taken to make sure whether the high or low metallicity branch is 
appropriate in each case.   
The uncertainty of the abundance is $\pm$0.2 dex.

\subsection{A sulphur abundance indicator}

The success of this empirical method of oxygen abundance determination has 
inspired a similar method for sulphur abundance determination. Since these
elements are iso-electronic, an expression corresponding to $\rm R_{23}$
can be made for sulphur, simply by replacing each oxygen line by the 
corresponding sulphur line. We define:\\[1ex]
\begin{equation}
\label{s23}
\rm S_{23} \equiv \frac{\sii ~\lambda \lambda 6716, ~6731 ~+ 
~\siii ~\lambda \lambda 9069, ~9531}{\hbeta}.
\end{equation}
\medskip
   
As two of the sulphur lines are in the far-red part of the spectrum, these
have only in the recent years become available, as detector sensibility 
in this spectral range has been improved. It might be more relevant to
normalize $\spunk$ to H$\alpha$, the nearest hydrogen Balmer line, 
than to $\hbeta$ lying further away in the blue range where it is more 
sensitive to extinction correction uncertainties. Our spectral resolution, 
however, is so poor that we can hardly deblend H$\alpha$ from the nearby 
[N\,{\sc ii}] $\lambda$6548 and [Ne\,{\sc ii}] $\lambda$6583 lines and thus  
we find it more reasonable to use $\hbeta$.
A calibration of $\rm S_{23}$ 
versus sulphur abundance should now enable sulphur abundance determinations
in \hii ~regions where the weak temperature sensitive line \siii 
~$\lambda$6312 is undetectable.

\begin{figure}[hbt]
\vspace{6.3cm}
\includegraphics{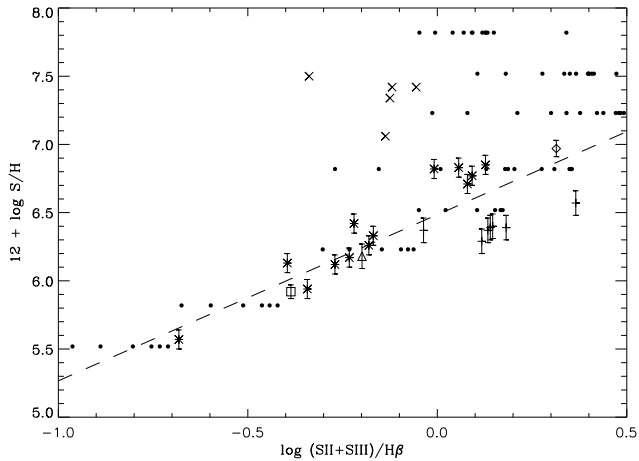}
\caption[S23cal]{\label{S23cal} Sulphur abundance as a 
function of~ $\spunk$.
$\times$~:~D\'\i az et al. (1991).
$\ast$~:~Garnett (1989).
$\triangle$~:~Garnett \& Kennicutt (1994).
$\Diamond$~:~Osterbrock et al. (1992).
+~:~Pastoriza et al. (1993).
$\Box$~:~Skillman et al. (1994).
$\bullet$~:~Theoretical models from Stasinska (1990), not included in the fit.
The dashed line is the fit given in Eq. (\ref{kalib}).}
\end{figure}

A search through literature for the data available for such a calibration
has been made. The resulting plot is shown in Fig. \ref{S23cal}.
All papers have measured values of emission line intensities of the constituent
lines of $\spunk$ and give from ionization model calculations the sulphur 
abundance.
The data points from D\'\i az et al. (1991) are from giant \hii ~regions 
in M51,
probably the spiral with the highest abundance known. Their position in
the plot suggests that they belong to a high metallicity branch, whereas
all other data points seem to fit nicely on a low metallicity branch.
It seems likely that only extremely high metallicity \hii ~regions will
be on the high metallicity branch. The majority of \hii ~regions will fit 
on the lower branch and thus only for this branch a fit is made.
A first order weighted fit gives:\\[1ex]
\begin{equation}
\label{kalib}
\log(\mathrm{S}/\mathrm{H})~+~\mathrm{12}~=~\mathrm{6.485}~+~
1.218\log(\spunk).
\end{equation}
\medskip

In addition a grid of model \hii ~regions from Stasinska (1990) has been
plotted in the diagram. All models have one ionizing star and both models
with uniform hydrogen number density 10 and 1000 cm$^{-3}$ have been included. 
The horizontal dispersion is mainly due to the effective
temperature of the ionizing star varying from 32500 to 55000 K. 
The theoretical models further consolidate the lower branch and suggest the
existence of a higher branch. The latter, however, has as yet a quite diffuse
appearance.

All papers used for this plot except Pastoriza et al. (1993) have taken
into account the possible contribution of S$^{3+}$ to the sulphur
abundance. All but one of their regions have so large O$^+$/O ratios 
($>$ 0.5)
that S$^{3+}$ is not likely to be present in significant amounts
(Garnett 1989). The rightmost of the datapoints in Fig. \ref{S23cal}
has considerably higher ionization (as it is a region in the nucleus),
and S$^+$ + S$^{2+}$ might not contribute more than 80 \% of the total
sulphur abundance.

\section{Observations}
The observations were carried out at the Danish 1.5-m telescope at ESO, La
Silla, Chile, where a facility to exploit multislit spectroscopy with the
Danish Faint Object Spectrograph and Camera (DFOSC) is under development. 
The
CCD available at the time of observation (Thompson 1k $\times$ 1k) had a field
size of $8\farcm6 \times
8\farcm6 $, large enough to include the main parts of the galaxy.
The aperture plates are manufactured on site from a hard plastic material with
a milling machine controlled by an instruction file generated on the basis of
well-sampled H$\alpha$-exposures obtained with the same instrument.

NGC 300 was observed during 2 nights in July 1995 (see Table \ref{log}). Two
different
aperture plates were used, both having 8 non-overlapping slitlets with widths
of $\simeq 2\farcs 5$. The number and lengths of the slitlets were governed
by the wish to maximize the number of \hrs~ observed in one exposure and the
need to have sufficiently sampling of the galaxy stellar continuum and the
local sky background, which is especially important in the near-IR range where
emission from atmospheric OH-lines rises strongly. Since some regions have
large angular sizes and some slitlets included two regions these criteria were
meet using lengths of 30--60\arcsec. One slitlet was positioned to contain pure
sky background for comparison, but the use of such a global sky spectrum gave
generally a poorer background subtraction, presumably due the low spectral
resolution (see below) and the displacement of the spectra in the dispersion
direction.

In total spectra of 14 \hrs~ were obtained with the two aperture plates; region
137A appears in the observations from both nights due to an initial
misidentification. With exposure times of 3000--7000 s the advantage of 
simultaneous observation of 5--10 \hrs~ compared to long-slit spectroscopy is
obvious.
Unfortunately the achieved spectral resolution was rather low, $\sim 25$ \AA{}
and $\sim 60$ \AA{} in the blue and red ranges respectively, due to a
malfunction of the best suited grisms. This means that \sii
~$\lambda\lambda$6716, 6731 must be deblended from the He{\sc\,i}
line at 6678 \AA{} and
the flux of weak Paschen lines  must be subtracted from
the \siii~ lines.

\begin{table*}
\caption[]{\label{log} Log of observations}
\begin{flushleft}
\begin{tabular}{lllc}
\noalign{\smallskip}
\hline
\noalign{\smallskip}
\hfil Date & \hfil \hr ID$^{\rm a}$ & \hfil Wavelengths (\AA ) & exp. time
 (s) \\
\noalign{\smallskip}
\hline
\noalign{\smallskip}
July 30 1995 & 45,61,77,79,109,118A,127,137A & 3650--6200 & 2700 \\
July 30 1995 & \hfil do. & 5300--10000 & 4500 \\
July 31 1995 & 53AB,76A,88+90,100,119A,137A,137C & 3650--6200 & 4800 \\
July 31 1995 & \hfil do. & 5300--10000 & 7200 \\
\noalign{\smallskip}
\hline
\noalign{\smallskip}
\end{tabular}
\begin{list}{}{}
\item[$^{\rm a}$] Identification of the individual \hrs~ follows the numbering
by Deharveng et al. (1988)
\end{list}
\end{flushleft}
\end{table*}

\section{Data reductions}
The multiaperture CCD frames were reduced by employing standard IRAF packages
for flat-fielding, extraction of one-dimensional spectra, calibration etc.
Since the slitlets cannot be milled with exactly the same widths special care
must be taken in the reduction process. Pixel-to-pixel variations are removed
by normalizing the spectrum from each slitlet using well-exposed dome
flat-fields made through the same instrumental setup. Long range variations over
the CCD
are accounted for by normalizing the collapsed one-dimensionel spectra from
each slitlet of a twilight sky exposure with the signal of the slitlet through
which the flux standard star exposures are taken. This last procedure is
similar to the illumination correction applied in long-slit spectroscopy.

The wavelength calibration was performed on the extracted one-dimensional
spectra
using He lamp exposures, both for the blue and red wavelength range, obtained
through the slitlet mask right after the science frames, while
the telescope was pointing in the same direction to minimize any effect of
flexure. Multiple exposures of the spectrophotometric standard star LTT 377
(Stone \& Baldwin 1983, Baldwin \& Stone 1984) from each night were used to
convert to absolute flux. The effect of atmospheric refraction (Fillipenko
1982) was negligible with the employed slit width, since both NGC 300 and LTT
377 were observed at low airmasses ranging between 1.03--1.08.
The blue and red spectra overlap in the region 5300--6200 \AA, which can be 
used to test the reliability of the flux calibration. Generally the 
deviation between the two ranges is less than 5 \% (see Figure \ref
{spec}) and the mean difference
for the apertures with high S/N is $\lid 2 \%$ for both nights, so no 
correction factor are applied when the two parts of the spectrum are merged.

In the near-IR spectral range careful measures were taken to remove the effect
of the atmospheric ${\rm H_2O}$ absorption band at $\lambda\lambda$8900--9800
using the procedure proposed by Osterbrock et al. (1990). After fitting
a smooth global continuum the line fluxes were measured by fitting a
Gaussian profile to the line, or two Gaussians if deblending from nearby
lines is relevant, and integrating under these profiles.

Finally the measured line fluxes are corrected for
interstellar extinction by determination of the visual extinction, $A_{\rm V}$,
using the line flux of H$\delta$, H$\gamma$, H$\beta$ and
H$\alpha$, and where possible the Paschen line P9, with the method
described by Petersen \& Gammelgaard (1996) assuming $T_{\rm e} = 10^4~{\rm K}$
and $n_{\rm e} = 10^2~{\rm cm^{-3}}$ for the giant \hrs. For
those \hrs~ where the
extinction determination does not seem to be reliable, values of $A_{\rm V}$
have been adopted from D'Odorico et al. (1983) and Webster \& Smith (1983) if
available, or as a simple average of the other regions.

The data reduction can be tested by comparing the ratio of the two near-IR
\siii ~lines with the theoretical ratio of 2.48 (Mendoza \& Zeippen 1982).
We find a mean value for $I(\lambda 9531) /  I(\lambda 9069)$ of 
2.78 $\pm$ 0.15, i.e. 12 \%
above the theoretical ratio. In fact this deviation is quite small
giving us faith that our correction for the atmospheric H$_2$O absorption band
and blending by OH emission lines has been successful in spite of the
low spectral resolution we were forced to use. This is probably due to the
fact that both
\siii$\lambda$ 9069 and \siii$\lambda$ 9531 are fortuitously positioned at 
wavelengths where atmospheric effects are relatively low as evident from the
plots in Osterbrock et al. (1990) and Osterbrock \& Martel (1992). We are
therefore confident
that the sulphur line fluxes can be assessed with sufficient accuracy for the
present study.

\section{Results}

\begin{figure}[hbt]
\vspace{6.3cm}
\includegraphics{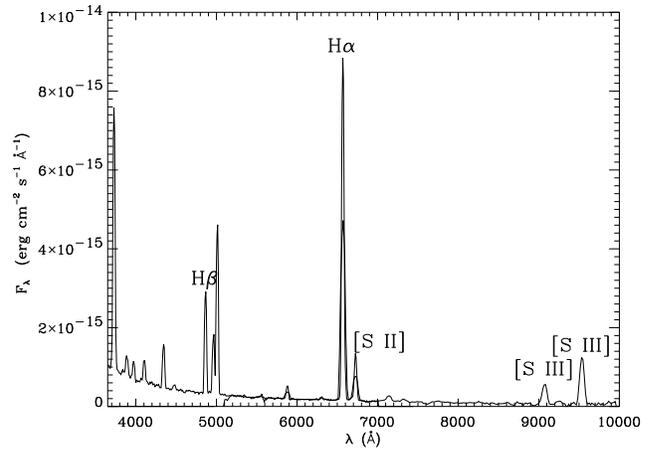}
\caption[spec]{\label{spec}  Combined spectrum of
the region 53A, B. Notice how smoothly the blue and red part of the
spectrum merge. In the overlapping region the difference in resolution
is obvious. The most important lines are labeled.}
\end{figure}

An example of the final reduced spectra is shown in Fig. \ref{spec}.
In the region 5300-6200 \AA , covered by both the blue and the red
exposure, the success of the flux calibration is demonstrated by
the closely coinciding spectral parts. 
In Table \ref{flux} the extinction and water absorption corrected line 
intensities relative to $\hbeta$ 
are listed along with the absolute $\hbeta$-flux and the extinction in V. 
The underlying Paschen hydrogen emission has been subtracted from the \siii 
~line fluxes by means of the theoretical line ratios from Hummer \& Storey 
(1987), assuming physical properties as mentioned above for the extinction
determination.

\begin{table*}[hbt]
\caption[]{\label{flux} Corrected line intensities }
\begin{flushleft}
\begin{tabular}{lllllllll}
\noalign{\smallskip}
\hline
\noalign{\smallskip} 
Region: & 45 \hfil & 53A, B \hfil & 61 \hfil & 76A \hfil & 77 \hfil & 79 \hfil 
& 88, 90 \hfil & 100 \hfil  \\
\noalign{\smallskip}
\hline
\noalign{\smallskip}
\oii ~$\lambda\lambda$3726, 3729 &  3.441 &  3.507 &  2.553 &  3.103 &  2.365 &
 3.018 &  4.366 &  2.034$^{\rm a}$ \\
H$\delta$ & 0.293$^{\rm a}$ & 0.294 &  &        &  &  & 0.419 &  \\
H$\gamma$ & 0.477 & 0.494 &  & 0.055$^{\rm a}$ &  &  &  &  \\
$\hbeta$ & 1.000 & 1.000 & 1.000 & 1.000 & 1.000 & 1.000 & 1.000 & 1.000$^{\rm a}$ \\
\oiii ~$\lambda$4959 &  0.105$^{\rm a}$ &  0.549 &  0.481 &  0.123$^{\rm a}$ 
&  0.464 &  0.380 &  0.254 &  0.221$^{\rm a}$ \\
\oiii ~$\lambda$5007 &  0.368$^{\rm a}$ &  1.580 &  1.503 &  0.297$^{\rm a}$ 
&  1.421 &  1.105 &  0.696 &  0.789$^{\rm a}$ \\
H$\alpha$ & 1.81$^{\rm a}$ & 2.25$^{\rm a}$ & & 2.28$^{\rm a}$ & & 
& 2.18$^{\rm a}$ & \\
\sii ~$\lambda\lambda$6716, 6731 &  0.623 &  0.344 &  0.292 &  0.478 &  0.376 
&  0.458 &  0.578 &  0.821$^{\rm a}$ \\
\siii ~$\lambda$9069 &  0.168$^{\rm a}$ &  0.217 &  0.214 &  0.285 &  0.164 &  0.175 &  0.169 &  0.557$^{\rm a}$ \\
P9 & 0.042$^{\rm a}$ & 0.038 &  & 0.038$^{\rm a}$ &  &  &  &  \\
\siii ~$\lambda$9531 &  0.331 &  0.484 &  0.516 &  0.860 &  0.348 &  0.370 
&  0.527 &  1.807$^{\rm a}$ \\
$-\log I(\hbeta)$ & 13.07 & 13.96 & 13.62 & 14.51 & 13.19 & 13.28 & 13.96 
& 13.43 \\
$A_{\rm V}$ &  1.04 &  0.91 &  0.87$^{\rm d}$ &  0.89 &  0.86$^{\rm b}$ 
&  1.01$^{\rm b}$ &  1.23 &  0.79$^{\rm c}$ \\
\end{tabular}
\end{flushleft}

\begin{flushleft}
\begin{tabular}{lllllllll}
\noalign{\smallskip}
\hline
\noalign{\smallskip} 
Region: & 109 \hfil  & 118A \hfil  & 119A \hfil & 127 \hfil & 137A \hfil 
& 137A \hfil &  137C \hfil  \\
\noalign{\smallskip} 
\hline
\noalign{\smallskip}
\oii ~$\lambda\lambda$3726, 3729 &  3.293 &  2.205 &  3.151 &  1.955 &  1.750 &
 2.094 &  2.523 \\
H$\delta$ &  & 0.226 & 0.254$^{\rm a}$ &  & 0.321$^{\rm a}$ &  &   \\
H$\gamma$ &  & 0.497 & 0.435 &  & 0.429 &  & 0.497$^{\rm a}$  \\
$\hbeta$ &  1.000 &  1.000 &  1.000 &  1.000 &  1.000 &  1.000 &  1.000 \\
\oiii ~$\lambda$4959 &  0.163$^{\rm a}$ &  0.597 &  0.348 &  0.747 &  0.862 
&  0.724 &  0.625 \\
\oiii ~$\lambda$5007 &  0.588$^{\rm a}$ &  1.771 &  1.041 &  2.293 &  2.583 
&  2.125 &  2.030 \\
H$\alpha$ & & & 2.80$^{\rm a}$ & & 2.18$^{\rm a}$ & & 2.31$^{\rm a}$ \\
\sii ~$\lambda\lambda$6716, 6731 & 0.582 & 0.291 & 0.467 & 0.302 & 0.162 
& 0.172 & 0.224 \\
\siii ~$\lambda$9069 & 0.101$^{\rm a}$ & 0.175 & 0.223 & 0.220$^{\rm a}$ 
& 0.270 & 0.198 & 0.114$^{\rm a}$ \\
P9 &  & 0.028$^{\rm a}$ & 0.024$^{\rm a}$ &  & 0.027$^{\rm a}$ &  
&  0.032$^{\rm a}$ \\
\siii ~$\lambda$9531 & 0.303 & 0.586 & 0.845 & 0.462 & 0.775 & 0.578 & 0.406 \\
$-\log I(\hbeta)$ & 13.42 & 13.17 & 13.74 & 13.70 & 13.14 & 13.79 & 13.18 \\
$A_{\rm V}$ &  1.22$^{\rm b}$ & 0.79 & 0.42 & 0.87$^{\rm d}$ & 0.67 & 0.67 & 1.70 \\
\noalign{\smallskip}
\hline
\noalign{\smallskip}
\end{tabular}
\end{flushleft}
\begin{list}{}{
}
\item[$^{\rm a}$] Uncertainty estimated to 30 \%. Elsewhere 15 \%.
\item[$^{\rm b}$] From D'Odorico et al. (1983).
\item[$^{\rm c}$] From Webster and Smith (1983).
\item[$^{\rm d}$] Average of the other regions.
\end{list}
\end{table*}

\begin{table*}[hbt]
\caption[]{\label{restab} Abundances }
\begin{flushleft}
\begin{tabular}{lllllllll}
\noalign{\smallskip}
\hline
\noalign{\smallskip} 
Region: & 45 \hfil & 53A, B \hfil & 61 \hfil & 76A \hfil & 77 \hfil & 79 \hfil 
& 88, 90 \hfil & 100 \hfil  \\
\noalign{\smallskip}
\hline
\noalign{\smallskip}
$\rho$/$\rho_o^{\rm a}$ & 0.27 & 0.39 & 0.32 & 0.09 & 0.35 & 0.35 & 0.37 & 0.07 \\
12 + log O/H$^{\rm b}$ & 8.91 & 8.71 & 8.84 &  8.96 &  8.87 &  8.84 & 8.75 & 9.02 \\
12 + log S/H$^{\rm c}$ & 6.55 &  6.51 &  6.50 & 6.74 & 6.42 & 6.49 & 6.61 & 7.10 \\
$-$log S/O$^{\rm c}$ & 2.37 & 2.20 & 2.34 & 2.22 & 2.45 & 2.36 & 2.13 & 1.92 \\
\end{tabular}
\end{flushleft}

\begin{flushleft}
\begin{tabular}{lllllllll}
\noalign{\smallskip}
\hline
\noalign{\smallskip} 
Region: & 109 \hfil  & 118A \hfil  & 119A \hfil & 127 \hfil & 137A \hfil 
& 137A \hfil &  137C \hfil  \\
\noalign{\smallskip} 
\hline
\noalign{\smallskip}
$\rho$/$\rho_o^{\rm a}$ & 0.17 & 0.29 & 0.34 & 0.30 & 0.40 & 0.40 & 0.42 \\
12 + log O/H$^{\rm b}$ & 8.90 & 8.84 & 8.84 & 8.79 & 8.76 & 8.79 & 8.76 \\
12 + log S/H$^{\rm c}$ & 6.48 & 6.51 & 6.71 & 6.48 & 6.58 & 6.46 & 6.33 \\
$-$log S/O$^{\rm c}$ & 2.42 & 2.32 & 2.13 & 2.31 & 2.18 & 2.34 & 2.44 \\
\noalign{\smallskip}
\hline
\noalign{\smallskip}
\end{tabular}
\end{flushleft}
\begin{list}{}{}
\item[$^{\rm a}$] From Deharveng et al. (1988).
\item[$^{\rm b}$] Uncertainty estimated to $\pm$ 0.2 dex.
\item[$^{\rm c}$] Typical uncertainty: $\pm$ 0.3 dex.
\end{list}
\end{table*}

An exact uncertainty calculation for the line intensities is very
hard to carry out due to the many sources of uncertainty in the
reduction steps and the deblending procedure. Consequently we have
assigned uncertainties to the lines following these qualitative 
guidelines inspired by Ryder (1995): when S/N is high and the lines easily
identified we assess a relative uncertainty of 15 \% and when
the lines could only just be distinguished from the noise 
we adopt a relative uncertainty of 30 \%. In the case of H$\alpha$
the uncertainty assigned is 30 \% due to problematic deblending from nearby
lines.

The agreement between the observations from the two nights might be
tested by comparing the two measurements of region 137A. The spectrum
from July 31st (listed first in Table \ref{flux}) has somewhat higher
resolution and the value of $A_{\rm v}$ derived from these data has 
been used for both spectra. 
The relative line intensities for \oii ~and \sii ~are lower the second 
night while the lines from the double-ionized stages are considerably
higher (20 \% for \oiii ~and 35 \% for \siii ).
This is more than the expected uncertainty and suggests that the
two slitlet positions do not coincide as we see a higher ionization
the second night and furthermore a much lower absolute $\hbeta$-intensity.

\begin{figure}[hbt]
\vspace{6.3cm}
\includegraphics{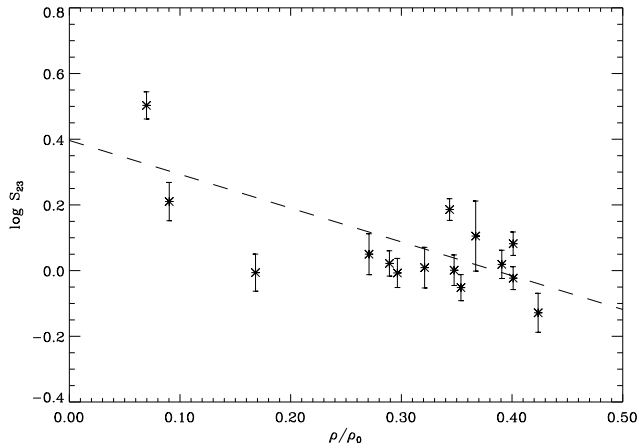}
\caption[s23gra]{\label{s23gra}  The variation of
the parameter $\spunk$ = (\sii +\siii )/$\hbeta$ across NGC 300. The abscissa 
gives the radial distance $\rho$ in units of the isophotal radius $\rho_o$. 
The dashed line is the best linear fit given in Eq. (\ref{gra}).}
\end{figure}

From the line intensities the parameter $\spunk$ is calculated 
according to Eq. (\ref{s23}) and
plotted versus radial distance from the center of the galaxy in Fig. 
\ref{s23gra}. The deprojected fractional isophotal radii $\rho$/$\rho_o$,
where the isophotal radius is the distance where the surface brightness 
equals 25 mag/arcsec$^2$, are taken from Deharveng et al. (1988).  For NGC 300
$\rho_o$ = 9\farcm 75. This normalization radius is less sensitive to 
selection effects, when comparing physical properties of spiral galaxies.
 The uncertainty calculation for $\spunk$ takes into
account the uncertainty of the extinction and the estimated  uncertainty
of the line fluxes. From Fig. \ref{s23gra} a gradient in $\spunk$ is 
evident. A linear weighted fit gives log $\spunk$ = 0.40 $\pm$ 0.09 $-$ 
(1.03 $\pm$ 0.29)$\rho$/$\rho_o$. By means of the relation in Eq.
(\ref{kalib}) this is converted to a sulphur abundance gradient:\\[1ex]
\begin{equation}
\label{gra}
\mathrm{12} + \log {\small \left(\frac{\mathrm{S}}{\mathrm{H}}\right)}
 = 6.97 \pm 0.14 - (\mathrm{1.25} \pm \mathrm{0.38}) ~\rho/\rho_o
\end{equation}
\medskip

For the rather high $\spunk$-values we find, one might wonder whether the
regions could belong to the (as yet indefinite) high metallicity branch in 
Fig. 
\ref{S23cal} in stead of the lower branch where our Eq. (\ref{kalib}) is valid.
However, since we find a decrease in $\spunk$ with radius, only the 
lower branch is consistent with a decreasing sulphur abundance gradient
which ensures us that our assumption is correct. A fit to the high
branch would be inversely proportional to the abundance and would
result in a sulphur abundance gradient increasing with radius in
contradiction with all expectations.

\begin{figure}[hbt]
\vspace{6.3cm}
\includegraphics{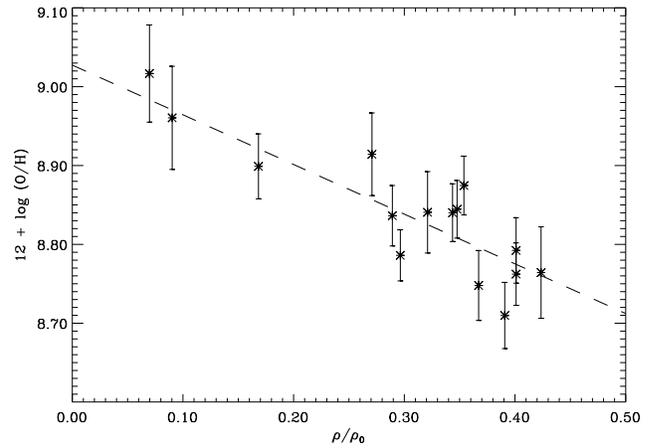}
\caption[ogra]{\label{ogra} The oxygen abundance gradient in NGC 300.
The abscissa gives the radial distance $\rho$ in units of the isophotal 
radius $\rho_o$. The dashed line is the best linear fit given in 
Table \ref{gradtab}.}
\end{figure}

From the observed oxygen line fluxes the parameter R$_{23}$ is calculated.
The calibration by ZKH of R$_{23}$ versus oxygen 
abundance is employed and the resulting plot of oxygen abundance
versus radius shows a clear trend of decreasing abundance with growing radius
(Fig. \ref{ogra}). In Table \ref{restab} the calculated abundances for each 
region are given as well as the isophotal radii.
The calculated abundance gradients are given in Table \ref{gradtab} together
with results for oxygen from previous work. There is a fine agreement between 
the results. 

Comparing the gradients in Figs. \ref{s23gra} and \ref{ogra} it is seen
that the oxygen abundance gradient is more well-determined than the one
for sulphur. This is most likely due to the much lower resolution of the 
red spectrum where all the sulphur lines are located as compared to
the blue spectrum where all the oxygen lines lie.

\begin{table*}[hbt]
\caption{\label{gradtab} Abundance gradients in NGC 300 }
\begin{flushleft}
\begin{tabular}{lr}
\noalign{\smallskip}
\hline
\noalign{\smallskip}
Source & \multicolumn{1}{c}{Abundance gradient} \\
\noalign{\smallskip}
\hline
\noalign{\smallskip}
This work & 12 + log $(\rm S / \rm H)$ = (6.97 $\pm$ 0.14) 
$-$ (1.25 $\pm$ 0.38) $\rho$/$\rho_o$ \\
This work & 12 + log (O/H) = (9.03 $\pm$ 0.04) 
$-$ (0.63 $\pm$ 0.13) $\rho$/$\rho_o$ \\
This work & log S/O = ($-$2.12 $\pm$ 0.11) $-$ 
(0.50 $\pm$ 0.35) $\rho$/$\rho_o$ \\
Deharveng et al. (1988) & 12 + log (O/H) = 
(8.95 $\pm$ 0.04) $-$ (0.63 $\pm$ 0.10) $\rho$/$\rho_o$ \\
Zaritsky et al. (1994)$^{\rm a}$ &  12 + log (O/H) = 
(8.97 $\pm$ 0.04) $-$ (0.61 $\pm$ 0.05) $\rho$/$\rho_o$ \\ 
\noalign{\smallskip}
\hline
\noalign{\smallskip}
\end{tabular}
\end{flushleft}
\begin{list}{}{
\itemsep=-3pt
}
\item[$^{\rm a}$] Based on the combined data from Webster and Smith (1983), 
Pagel et al. (1979) and Deharveng et al. (1988).
\end{list}
\end{table*}

\begin{figure}[hbt]
\vspace{6.3cm}
\includegraphics{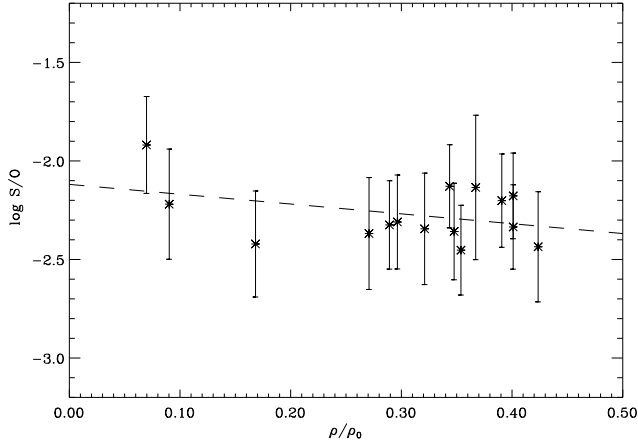}
\caption[sogra]{\label{sogra} log S/O plotted vs. fractional isophotal
radius $\rho$/$\rho_o$. The dashed line represents the best linear fit, 
see Table \ref{gradtab}. }
\end{figure}

As the sulphur abundance has a steeper slope than the oxygen abundance we
find a slight decrease in S/O as the radius increases as can be seen in 
Fig. \ref{sogra}. The value of the gradient is given in Table \ref{gradtab}. 

From Fig. \ref{s23gra} it is apparent that the value of the sulphur abundance
gradient is quite dependent on the innermost datapoint. If this region
(\# 100) is omitted the gradient is drastically reduced to $-$0.43 $\pm$ 0.37
dex/$\rho_o$, indicating that the uncertainty of the gradient might very well
be greater than that quoted in Table \ref{gradtab}.
We do not, however, especially mistrust our observations of this 
region. Even though it has the weakest emission of our sample the S/N of 
the spectrum is acceptable. 
To firmly establish the value of this gradient observations of more
regions at large radial distances as well as at small distances are needed
(preferably at a higher resolution).

The same sensitivity to this innermost datapoint is displayed by the 
gradient in Fig. \ref{sogra}.
When the S/O gradient is calculated with region 100 omitted the result is
$+$0.12 $\pm$ 0.36 dex/$\rho_o$, i.e. a radial gradient in S/O is no longer
present. 

If a relation between log S/O and oxygen abundance could be established
a powerful tool to test galactic chemical evolution models would be at hand.
Sulphur is produced in stars with M $>$ 10 M$_{\sun}$ and the ratio of sulphur 
to oxygen produced is greatest in stars with masses in the range 12--20 solar
masses. Variations in log S/O vs. metallicity could thus reflect variations in 
the IMF. Garnett (1989). Investigations of this relationship has been made by
Garnett (1989) and D\'\i az et al. (1991). The former find no evidence for
a relation while the latter investigates more metal rich regions and find
a decrease at high metallicities.
Our data confirm the trend demonstrated  by D\'\i az et al. (1991)
of a decrease in log S/O at high oxygen abundance. We
do, however, find quite low S/O-values in comparison with their 
collection of data, but keeping the rather large uncertainties
(including both uncertainties of the fluxes and the fit in Eq. (\ref{kalib}))
in mind, the discrepancy might not be significant.

\subsection{The softness and ionization parameters}
Some of the physical parameters determining the \hr~ spectra are the ionization
parameter $u$ and the ionizing spectrum characterized by the mean effective
temperature $T_{\star}$ of the ionizing
stars. V\'\i lchez \& Pagel (1988, hereafter VP) introduced the parameter $\eta =
({\rm O}^+ /
{\rm O}^{2+})/({\rm S}^{2+} / {\rm S}^+)$, which was shown to have a one-to-one
relation with $T_{\star}$, as an indicator of the radiation softness. It has
the important advantages of being insentive to density and virtually
independent of $u$ and $T_{\rm e}$. The actual 
relationship with $T_{\star}$ depend on stellar model atmosphere calculation. 

As discussed by Garnett (1989) it is still premature to make any quantitative
assessment of $T_{\star}$ from $\eta$ and more detailed non-LTE models of
hot stars are required to derive absolute values of the effective
temperature of the ionizing sources.
Nevertheless, he argues that $\eta$ can still be used as a sequencing
parameter to examine the relative behaviour of $T_{\star}$, and we shall use
it to investigate the gross proporties of $T_{\star}$ in NGC 300. When the 
near-IR
\siii~lines are available $\eta$ is easily found from the observable line
ratio quotient (VP):
\be
\eta\prime = \frac{\oii~\lambda\lambda3726,3729}{\oiii~\lambda\lambda
4959,5007} \left/ \frac{\sii~\lambda\lambda6717,6731}{\siii~\lambda\lambda
9069,9531} \right.
\ee
through the relation
\be
\log \eta = \eta\prime + 0.14 / t + 0.16,
\ee
where $t = T_{\rm e} / 10^4 {\rm K}$. We used the value $t=1.0$ for all \hrs~
rather than acquiring $T_{\rm e}$ from various sources,
since any variation of $t$ in the range 0.7--1.4 will only result in a change in
$\log \eta$ of 0.1 dex (VP), less than the observational
errors, and the few regions that have $T_{\rm e}$ determined in the literature
(Webster \& Smith 1983; Pagel et al. 1979)
are all in the range 8000--11000 K.
\begin{figure}[hbt]
\vspace{6.3cm}
\includegraphics{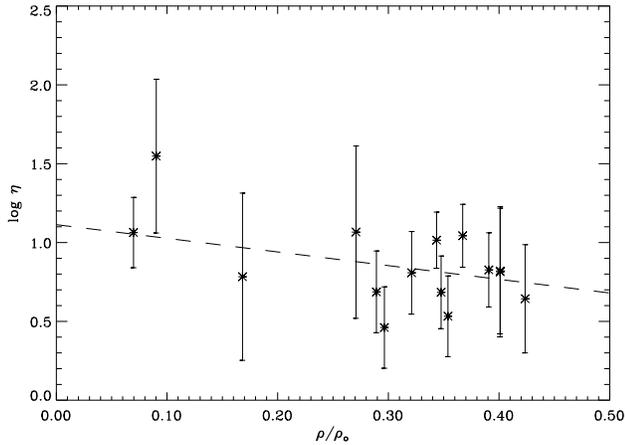}
\caption[s23gra]{\label{eta} The softness parameter $\eta$ plotted vs. the
fractional isophotal radius $\rho / \rho_o$. The dashed line represents
the best linear fit.}
\end{figure}

In Fig. \ref{eta} we plot the
deduced $\eta$-parameter for the \hrs~ in NGC 300 versus fractional isophotal 
radius which
can be compared to Fig. 5 of ZKH. Our data points fall in the bulk of their
measurements for seven spiral galaxies between $\rho / \rho_o$ = 0--0.5, thus 
falling below the original fit to M33 and M101 by VP. Given the uncertainties
in all works discussed here this might not be
significant, though. We confirm for NGC 300 the general trend of decreasing
$\log \eta$ with increasing radius.

VP interpret a correlation of $\log \eta$ with oxygen abundance
in the sense that at low oxygen abundance stars tend to be hotter,
as a result of an IMF depending on abundance,
possibly with some scatter related to evolution. This implies that for galaxies
with steep abundance gradient one expect to find gradients in $T_{\star}$ and
likewise in $\eta$, while low-abundance galaxies with no
destinctive gradient only will show a scatter due to mixing of evolutinary
stages. 
When compared to the plot by ZKH NGC~300 indeed seems to have an intermediate
slope of the $\log \eta$-gradient resembling maybe that of NGC~2903 and
NGC~4559, which have similar oxygen gradients and shallower than the fit for
M33 and M101, which also have the steepest oxygen gradients.

The ionization parameter may be calculated from the ratio of the \sii~ and
\siii~ lines through the relationship given by D\'\i az et al. (1991)
\be
\log u = -1.68\, \log\left( \frac{\sii~\lambda\lambda6717,6731}
{\siii~\lambda\lambda 9069,9531} \right) - 2.99, 
\ee
valid for log(\sii / \siii) $\simeq$ -0.5 -- 1.0.
The values of log~$u$ fall in the range $-1.6~{\rm to}~-3.3$ with most of the
regions scattered around $-2.5$, thus justifying the use of Eq. (6). No trend
with isophotal radius was seen as was also the 
case for the seven spiral galaxies where $u$ has been investigated by ZKH.

\section{Conclusion}
The observations of NGC 300 have demonstrated the advantage of the 
MOS facility for spectroscopy of extragalactic \hrs , where
15--20 regions
distributed over a wide field can be included in only two exposures. In this 
setup the individual slitlets can be made sufficiently long to enable a good
sampling of the local sky background, which is necessary to obtain precise
spectra at all wavelengths in the range 3650--10000 \AA .

We have successfully applied a new sulphur abundance determination
scheme based on the strong \sii{} and \siii{} emission lines to \hrs~in the 
spiral galaxy NGC 300 in spite of the very poor resolution attainable
at the time of the observations. 
The regions exhibit a clear trend of decresing sulphur abundance with
increasing galactocentric distance with a gradient of $- 1.25 \pm 0.38\,
{\rm dex} / \rho_o$, as well as a distinct oxygen abundance gradient.
It is assuring that the latter is consistent with previous work (Table 
\ref{gradtab}). 
Quite low S/O-values for NGC 300 are derived in support of the
decrease found by D\'\i az et al. (1991) at high metallicity
but any certain conclusions on this issue must await more data
from a larger sample of galaxies.
The radiation softness parameter seems to have a gradient with a slope
equal to galaxies of similar oxygen abundance gradients.

The sulphur abundance gradient found in this project
could be improved by including regions with $\rho / \rho_o > 0.5$,
preferably at distances at or beyond $\rho_o$.
With the large field size and sensitivity of the CCD's now used,
multislit spectroscopy will undoubtedly be a
powerful tool for abundance studies in spiral galaxies.

In the future we will apply our empirical sulphur abundance method to 
\hrs~in other spirals. Hopefully further detailed studies of sulphur
abundance in bright \hii ~regions will lead to an improvement
of the calibration of the suggested empirical sulphur abundance indicator,
$\spunk$.
\begin{acknowledgements}
We highly appreciate the advices concerning the MOS facility
at the Danish 1.5-m telescope at ESO, La Silla by Michael I. Andersen,
without whom the multislit observations could not have been accomplished.
This project was supported by the Danish Board for Astronomical Research.
\end{acknowledgements}

\end{document}